\documentclass[letterpaper, 10pt, conference]{ieeeconf}
\IEEEoverridecommandlockouts	

\usepackage{multirow}
\usepackage{lipsum}
\usepackage{amsmath}
\usepackage{amssymb}
\usepackage{subcaption}
\usepackage{graphicx}
\graphicspath{{./figures/}}
\usepackage{glossaries}
\usepackage[table,xcdraw]{xcolor}
\usepackage[ruled,vlined]{algorithm2e}
\usepackage[final]{listings}
\usepackage{longtable}
\usepackage[normalem]{ulem}
\usepackage{tikz,xcolor,hyperref}
\definecolor{lime}{HTML}{A6CE39}
\DeclareRobustCommand{\orcidicon}{
	\begin{tikzpicture}
	\draw[lime, fill=lime] (0,0) 
	circle [radius=0.16] 
	node[white] {{\fontfamily{qag}\selectfont \tiny ID}};
	\draw[white, fill=white] (-0.0625,0.095) 
	circle [radius=0.007];
	\end{tikzpicture}
	\hspace{-2mm}
}
\foreach \x in {A, ..., Z}{\expandafter\xdef\csname orcid\x\endcsname{\noexpand\href{https://orcid.org/\csname orcidauthor\x\endcsname}
			{\noexpand\orcidicon}}
}

\definecolor{mygray}{rgb}{0.5,0.5,0.4}
\definecolor{mygreen}{rgb}{0.2,0.5,0.2}
\definecolor{myblue}{rgb}{0.2,0.2,0.9}
\definecolor{mykwdclr}{rgb}{0.2,0.6,0.6}

\lstdefinestyle{mybase} {
	language=[Sharp]C,
	breaklines=true,
	showstringspaces=false,
	basicstyle=\small\tt,
	frame=single,
    numbers=left,
    numbersep=8pt,
	numberstyle=\tiny\color{mykwdclr},
	keywordstyle=\color{myblue},
	keywords=[2]{Mathf,Laser_Scanner,MonoBehaviour,Vector3,RaycastHit,Debug,Color,Physics}, 
	keywordstyle=[2]\color{mykwdclr}, 
	commentstyle=\color{mygreen}\ttfamily
}

\lstdefinestyle{mycsh} {
	style=mybase
}

\newacronym{ITS}{ITS}{Intelligent Transportation Systems}
\newacronym{ROS}{ROS}{Robot Operating System}
\newacronym{SUMO}{SUMO}{Simulation of Urban MObility}
\newacronym{URDF}{URDF}{Unified Robot Description Format}
\newacronym{SLAM}{SLAM}{Simultanous Localization and Mapping}
\newacronym{JSON}{JSON}{JavaScript Object Notation}
\newacronym{LDS}{LDS}{Laser Distance Sensor}
\newacronym{LIDAR}{LIDAR}{Light Detection And Ranging}

\title{\LARGE \bf Real-World Evaluation of the Impact of Automated Driving System Technology on Driver Gaze Behavior, Reaction Time and Trust}

\author{Walter Morales-Alvarez$^{1}$\orcidD \emph{Student Member, IEEE}, Mohamed Marouf $^{2}$, \\ Hadj. Hamma Tadjine \emph{Senior Member, IEEE} $^{3}$ and Cristina Olaverri-Monreal$^{1}$\orcidE{} \emph{Senior Member, IEEE}%
\thanks{$^1$ Johannes Kepler University Linz; Chair Sustainable Transport Logistics 4.0, Altenberger Straße 69, 4040 Linz, Austria.
	\texttt{\{walter.morales\_alvarez, cristina.olaverri-monreal\}@jku.at}}%
	\thanks{$^2$ IAV France S.A.S.U., 4 Rue Guynemer, 78280 Guyancourt, France.
	\texttt{mohamed.marouf@iav.de}}%
		\thanks{$^3$ IAV GmbH Entwicklungszentrum, Carnotstraße 1,  10587 Berlin, Germany.
	\texttt{hadj.hamma.tadjine@iav.de}}%
}

\begin{document}

\maketitle
\thispagestyle{empty}
\pagestyle{empty}
\captionsetup[figure]{name={Fig.},labelsep=period}
\begin{abstract}
Recent developments in advanced driving assistance systems (ADAS) that rely on some level of autonomy have led the automobile industry and research community to investigate the impact they might have on driving performance. However, most of the research performed so far is based on simulated environments. In this study we investigated the behavior of drivers in a vehicle with automated driving system (ADS) capabilities in a real life driving scenario. We analyzed their response to a take over request (TOR) at two different driving speeds while being engaged in non-               driving-related tasks (NDRT). Results from the performed experiments showed that driver reaction time to a TOR, gaze behavior and self-reported trust in automation were affected by the type of NDRT being concurrently performed and driver reaction time and gaze behavior additionally depended on the driving or vehicle speed at the time of TOR.     

\end{abstract}

\section{Introduction}
\label{sec:introduction}
Vehicles that are able to perform the Driving Dynamic Task (DDT) with certain automation are already available on     the market. 
The introduction of these automated functions in vehicles is promising as the absence of human intervention in the control of vehicles will increase road safety~\cite{olaverri2016human}. However, automation in a vehicular environment represents a complex problem, as it might lead to a reduction in driver situational awareness~\cite{olaverri2016autonomous,olaverri2017road} and      make it necessary to monitor driver behavior in case manual control needs to reinstated~\cite{Olaverri-Monreal2020}.

In this context, the introduction of driving monitoring systems (DMS) allows automated driving systems (ADS) to estimate driver status and behavior and convey the information according to the physical and mental state of the driver. 

Automated systems are not yet able to respond to all situations that may occur on the road~\cite{morales2020automated, ntsb2020}. Conditional automation (SAE level 3) is the first level where ADS is capable of performing the entire driving dynamic task, allowing users to perform non-driving related tasks (NDRT). When the system exceeds its operational design domain (ODD) or there is an emergency situation, the vehicle issues a take over request (TOR) indicating to the drivers that they must take control of the vehicle. In certain situations, this transition can be complex, as the driver might be engaged in other tasks and / or not paying attention to the road. In these cases, the drivers must be aware of the road situation to be able to safely take back the control of the vehicle.  

This situational awareness decreases, as previously mentioned, if the drivers are involved in NDRT, as multiple tasks compete for the driver's attention and augment the cognitive workload~\cite{RUSCIO2017105, Blatt1994}. 
The use of smartphones has dramatically increased in the last years, representing a major cause of road accidents. Currently, smartphones are designed to maximize the amount of attention that users devote to them. Even if their use while driving is banned in most countries, they continue to be used, and the introduction of vehicles equipped with conditional automation or level 3 will promote their use. Therefore, it is crucial to evaluate the consequences.

In this work we examined the impact of automation on driver behavior, specifically analyzing the effect of driving/vehicle speed and NDRT on the drivers ability to regain control of a vehicle controlled by an ADS. To this end we defined speed and NDRT as the independent variables that we modified to measure the dependent variables, which were related to reaction time and gaze behavior. We additionally investigated the level of confidence and trust in the automation. To this end we formulated the following null and alternative hypotheses for the cases below: 

\subsection{Impact of secondary tasks (NDRT) on driver gaze behavior}
\label{subsection:ndstgb}

\begin{itemize}
\item H0: Secondary tasks do not have any effect on the amount of time the driver directs their gaze away from the road.
\item H1: Secondary tasks increase the amount of time the driver directs their gaze away from the road. 
\end{itemize} 

\subsection{Relationship between driving speed and driver gaze behavior} 
\label{subsection:sgb}

\begin{itemize}
\item H0: The driving speed does not have any effect on the amount of time the driver directs their gaze away from the road.
\item H1: Increased driving speed decreases the amount of time the driver directs their gaze away from the road. 
\end{itemize} 

\subsection{Effect of the secondary task (NDRT) on driver reaction time}
\label{subsection:srt}

\begin{itemize}
\item H0: Secondary tasks (NDRT) do not have any effect on driver reaction time.
\item H1: Secondary tasks (NDRT) increase driver reaction time.
\end{itemize} 

\subsection{Effect of the vehicle speed on driver reaction time}
\label{subsection:vrt}

\begin{itemize}
\item H0: Vehicle's speed do not have any effect on driver reaction time.
\item H1: Vehicle's speed affects driver reaction time.
\end{itemize} 

\color{black}

\subsection{Influence of driving speed on self-reported trust in the automation} 
\label{subsection:sst}

\begin{itemize}
\item H0: Vehicle speed does not affect driver self-reported trust in the automation when performing a secondary task and in the baseline condition.
\item H1: When performing NDRT or in the baseline condition, driver self-reported trust in the automation decreases as vehicle speed increases.
\end{itemize} 

\subsection{Influence of performed tasks on self-reported trust in the automation} 
\label{subsection:ssttask}
\begin{itemize}
\item H0: Performing NDRT does not affect driver self-reported trust in the automation.
\item H1: When performing NDRT, driver self-reported trust in the automation decreases.
\end{itemize} 

The  remainder  of  the  paper  is  organized  as  follows: the next  section  explains related  studies in  the  field;  section~\ref{sec:fieldtest} details the field test setup; section~\ref{sec:evaluation}  presents  the  method  used to acquire and process the data collected; sections~\ref{sec:results} and~\ref{sec:findings} present and interpret the obtained results respectively; the final section discusses and concludes the present study.

\section{Related Work}
\label{sec:relatedwork}

The factors that influence drivers during TOR situations have been studied throughout the years in several works. For example, external factors such as a high traffic density of surrounding vehicles lead to a proportional increase in lateral acceleration and decrease in time to collision~\cite{Gold2016TakingSituations,Radlmayr2014,8594655,gold2018modeling, du2020evaluating}.  

As previously mentioned, a TOR is triggered when the road conditions do not allow for the automation to work as expected because the automated system has exceeded its ODD. This is the case, for example, in curvy roads driven at high speed~\cite{Martens1997THEEO}. Several studies have investigated driver reactions in this scenario, the work in~\cite{9294751} showing that the road features affect the driver’s level of visual attention and trust in the automation. Additional research indicated that the existence of curves increased lateral acceleration and reaction time to respond to a TOR~\cite{Neukum2014}. Furthermore, the authors in~\cite{Brandenburg2019Take-overConditions} argued that drivers tend to make abrupt deceleration maneuvers after regaining the control of the vehicle when entering a curve.

In line with this, the relationship between urgency, road geometry and driver reaction time after a TOR has also been a subject of research in the last years, for example in the work in~\cite{Borojeni2018FeelVehicles}, which shows that the reaction time to a TOR increased when urgency cues were conveyed in a curvy scenario.

Another important factor that affects the process of regaining control of the vehicle is the performance of NDRT~\cite{9294751, du2020evaluating, radlmayr2014a, zeeb2016take, wandtner2018effects}. Furthermore, tasks that involve a high cognitive load \cite{du2020evaluating} increase driver reaction time to a TOR compared to low cognitive load tasks.

This was shown to be the case in the simulated environment study ~\cite{gold2013take}, where an estimated 7 seconds was necessary to take over control of the vehicle.
Other studies that reported the take over reaction time (TOrt) in different scenarios showed that TOrt stayed fairly consistent,  between 2 and 3.5 seconds~\cite{Eriksson2017}. 
However, experiments performed under real driving conditions showed that drivers needed less than one second to take back the control of the vehicle, this time varying depending on the road conditions and tasks performed~\cite{9294751}.

How to promote trust in autonomous vehicles has also been the topic of research in a variety of works~\cite{Olaverri-Monreal2020, allamehzadeh2016automatic}. In this context, acceptance of the technology depends on      whether the mental models of users and systems match \cite{verberne2012trust, olaverri2014capturing}. Furthermore, \cite{hussein2016p2v, petersen2019situational, beller2013improving} indicated that trust in automation increased when information regarding the environment or warning cues regarding system failures were conveyed. Related to this, a higher automatism of vehicles might lead to a reduction in driver situational awareness as a consequence of the hypovigilance~\cite{capalar2017hypovigilance} that results from overtrusting the systems~\cite{xiong2012use}.

Even though the most convenient way of transferring control from ADS-based vehicles to the driver has been widely addressed in the literature, most of the existing results have been obtained from experiments in virtual environments. Studies that assess driver response to TOR in real road scenarios are very scarce. In this work we build on the research presented in~\cite{9294751}, in which experiments were conducted in a real life driving scenario and thereby contribute to the scientific community by investigating the impact of automation on driver behavior under several speed and NDRT conditions that have not been previously considered in real driving conditions. 

\section{Field Test Description}
\label{sec:fieldtest}

\subsection{Participants}

Our sample consisted of 14 participants with an average age of 27.06 (SD = 10.15) and who did not have previous experience with ADS. The participants in the experiments were recruited among employees from the ÖAMTC Driving Technology Centre in Marchtrenk, Upper Austria and Teesdorf, Low Austria, from the Johannes Kepler University in Linz and the University of Applied Sciences Technikum Wien in Vienna.  

\subsection{Apparatus and experiment scenario}

\begin{figure}
	\centering
	\includegraphics[width=0.48\textwidth]{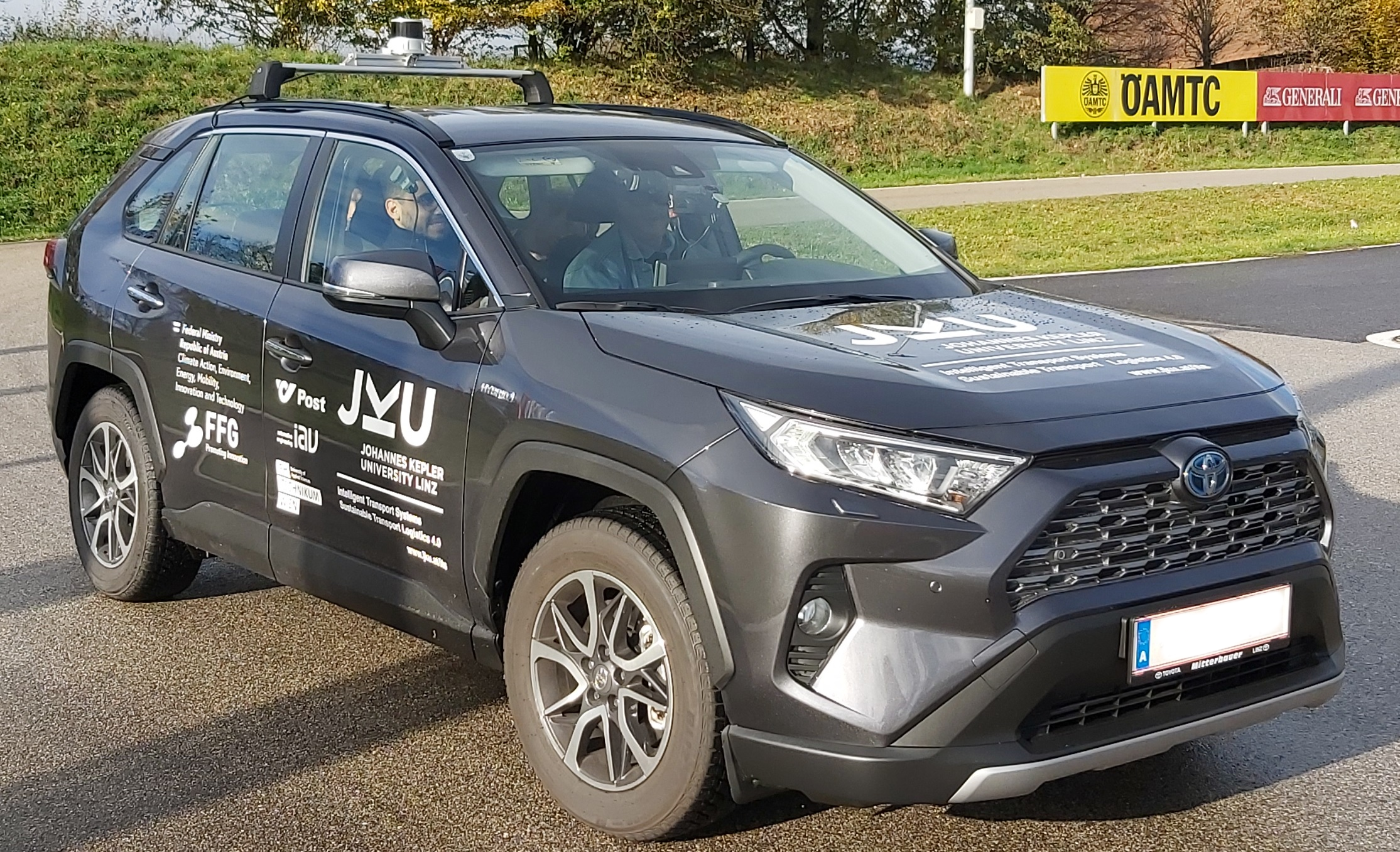}
	\caption{ Vehicle with automated capabilities that was used to perform the tests. }
	\label{fig:apparatus}
\end{figure}

The experiments were performed using a Toyota RAV4 2020  that was equipped with several sensors for monitoring the environment and / or the driver as well as external and internal cameras~(Figure~\ref{fig:apparatus}). We equipped the vehicle with an Openpilot system that was connected to the car through the OBD-II. Openpilot is an open source semi-ADS designed by the company Comma.ai. It is installed in a device called Comma Two and provides vehicles with automated features such as ACC, lane keeping and lane change~\cite{openpilot}. Using a controller area network (CAN) database written by the company, the device is able to control the vehicle's actuators through the vehicle's ADASs. In this way, the system requires only a road lane and a target speed to drive, without the driver being required to steer the vehicle or control it through the acceleration and / or brake pedals.

The experiments were conducted at the Austrian Automobile, Motorcycle and Touring Club (ÖAMTC) Driving Technology Centre in Marchtrenk and Teesdorf, in Upper and Lower Austria, respectively. 
To perform the experiments described in this work we used tracks with the same characteristics in both locations that consisted of a straight 645 meter stretch of road. For safety reasons, no other vehicle was on the road. Additionally, a supervisor in the passenger seat monitored the functioning of the vehicle's systems and controlled the vehicle if the driver did not respond to a triggered TOR. This was necessary to guarantee safety as a TOR corresponded in many cases to a vehicle operation outside its ODD.

\begin{figure}
	\centering
	\includegraphics[width=0.48\textwidth]{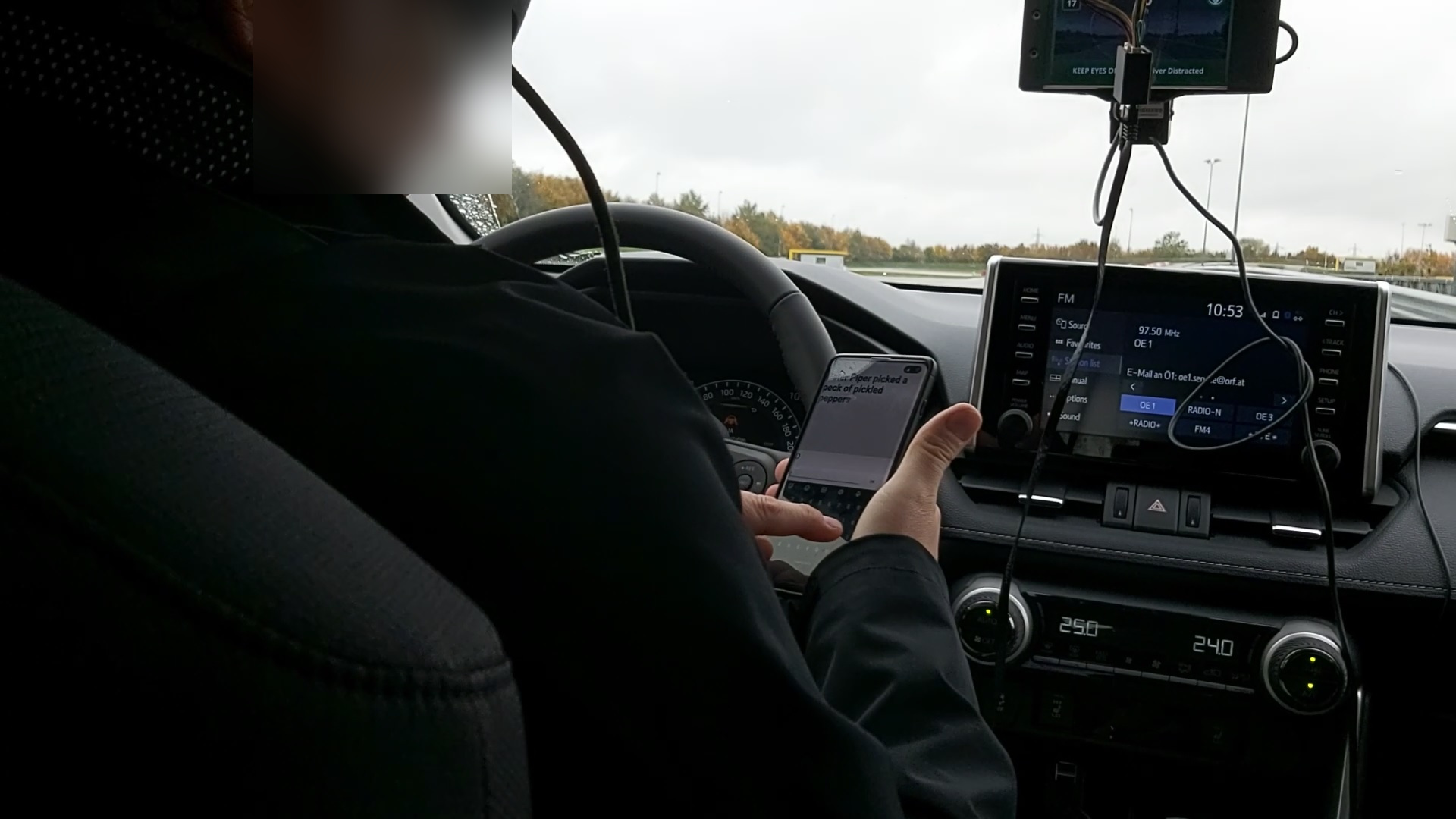}
	\caption{ Participant performing the visual and manual secondary tasks while the vehicle was controlled by the automated system. }
	\label{fig:participant}
\end{figure}

The NDRT (also referred to in this work as ``secondary tasks'') were performed using a 6.4 inch display Samsung S10 mobile phone that the participants held while performing the tasks (Figure~\ref{fig:participant}). To this end, we implemented one application in Unity to produce the tasks on the Samsung. Furthermore, the participants were fitted with a pair of VPS19 eye tracking glasses that were developed and sponsored by the company Viewpoint system GmbH~\cite{vps}, and which allowed the assessment of the driver’s visual point of fixation.

\subsection{Non- driving-related secondary tasks}

The conducted experiments involved the following conditions and tasks to provide cognitive load and manual/visual occupation. The order of the tasks and speed scenarios was alternated for each driver to avoid bias. The duration of the tasks during the automation was the same for all the participants when they were asked to regain the control of the vehicle.

\begin{itemize}
    \item \textbf{Baseline}: ADS was engaged without a secondary task being performed.
    \item \textbf{Visual}: The tasks consisted of performing the Stroop Color and Word Test (SCWT)~\cite{jensen1966stroop}. In this task the screen of the mobile phone displayed a color name in text of a different color (e.g the word ``red'' was written in yellow text). The participant needed to speak out loud the color in which the word was written.
    \item \textbf{Manual}: To replicate a real life situation in which drivers search for items in a bag, participants had to retrieve three M8x8 screws that were inside a bag filled with chocolate balls of 1 cm radius. 
    \item \textbf{Visual and Manual}: Participants were required to write a given text backwards on the mobile phone provided.
\end{itemize}

\subsection{Experimental setup and procedure}

The experiment consisted of driving a car in the described scenario. The vehicle was initially stopped. After being activated, OpenPilot initiated driving and the system performed the complete DDT. When the given speeds that had been previously defined (30 and 50 km/h in each round) were reached, the participants in the experiment performed the three secondary tasks described in the previous subsection until the system issued a TOR that demanded situation awareness and a positive control of the vehicle.

The TOR occurred 30 seconds after the OpenPilot was engaged and consisted of an auditory signal issued by an additional mobile phone.
Prior to the tests, the participants were provided with a pair of eye tracking glasses, which they had to calibrate following the instructions of the supervisor. Before starting each experiment the participants were instructed how to activate and deactivate the OpenPilot system. Releasing the brake pedal and pressing the set button in the steering wheel engaged the OpenPilot system.  OpenPilot was disengaged by putting the hands on the steering wheel since control of the vehicle was required when the TOR was issued. Alternatively it could also be disengaged by pressing the gas or brake pedal.
This setup was chosen to replicate the same conditions (baseline and tasks condition duration) for all the participants and guarantee a repeatable experiment. 

A training session without a secondary task was performed before the test so that the participants could get accustomed to the vehicle's system. During the driving experiment, each participant drove a total time of 30 minutes during which the three different tasks were performed, without break between segments. At the end, each participant completed a post-task questionnaire to collect subjective data regarding the experiment.


\section{Data Collection and Evaluation}
\label{sec:evaluation}

In order to test the hypotheses defined in section~\ref{sec:introduction}, we focused on the time that drivers needed to take back control of the vehicle, gaze behavior and self-reported trust in the automation at different speeds and depending on the secondary tasks performed.

\subsection{Data acquisition}
To measure the impact of automation on driver behavior, we first acquired the corresponding data through the following means: internal camera, cross framework that combined the Robot Operating System (ROS) and Openpilot for the reaction time and eye tracking using  Viewpointsystem’s own framework. In addition, through a post task questionnaire with a Likert scale from 0 to 5 (5 being the highest score) we collected data regarding the level of trust on the automation depending on the tasks performed and vehicle velocity. Additional questions related
to demographic information and a field to enter comments
completed the survey. From this data we determined the following dependent variables:

\begin{itemize}
	\item \textbf{Frequency of eyes from task to road (Fr)}: 
	Defined as the frequency that drivers looked from the secondary task to the road or number of switches.
	\item \textbf{Point of visual fixation {(PF)}:} Defined as the exact point at which the drivers were looking to  perform the secondary task.
	\item \textbf{Average time of eyes on road {(AVR)}:} Defined as the average time that drivers were looking at the road.
	\item \textbf{Average time of eyes on task {(AVT)}:} Defined as the average time that the drivers were looking to perform the secondary task. 
	\item \textbf{Reaction time {(RT)}:} Defined as the time required      for the participant to take      control of the  vehicle, starting from the moment in which a TOR was triggered. 
	\item \textbf{Self-reported trust {(ST)}:} Defined as the perceived level of trust in the automation.
\end{itemize}

\subsection{Data processing}
To process the data for future analysis, we first performed a synchronization process between the data obtained by OpenPilot and the sensors installed on the vehicle that were controlled by the ROS framework. For the synchronization, initially a ROS node was programmed to translate the data recorded by OpenPilot that were transmitted through the Cereal library developed by comma.ai. Cereal is a messaging specification for robotics systems, which uses the ZeroMQ messaging library as Back-End. The implemented ROS node then accessed the messages sent by OpenPilot and bundled them, allowing them to       be sent by ROS topics as predefined messages that were then synchronized based on their timestamp. 
In order to synchronize the data obtained from the eye tracking glasses and the ROS data, the information from the glasses was first extracted using a specific software of the company that made it possible to obtain a video as well as processed information per frame. Using the data frequency and start timestamp a new timestamp was generated for each frame. Finally, a node was created to send the extracted videos with the relevant information and the generated timestamps that were synchronized with the sensor data.



After the data synchronization, the reaction time was obtained through a node that was in charge of extracting the elapsed time between the frame when the TOR was issued and the frame in which OpenPilot was disengaged by the driver. For this, the node took into account the allocated time of 30 seconds for the task performance and the status of the vehicle and system obtained by OpenPilot.

The gaze was obtained by extracting the timestamp at which the visual point of fixation was located in the field of vision of the road and the time at which the visual point of fixation moved between the field of vision required to perform the task and the road. For this purpose, a ROS node was used to manually mark the frames with the participant’s visual focus and was able to extract and subtract the timestamps of each frame.

\subsection{Statistical Analysis}

To determine whether a statistically significant relationship existed between the dependent variables and the independent variables of secondary tasks and the vehicle speed, we performed a  One Way ANOVA with a Bonferroni multiple post-hoc comparison. We additionally conducted a non parametric Wilcoxon signed-rank  test for dependent samples (as data were obtained from repeated measurements) to investigate statistical significant results regarding the effect of the performed tasks and vehicle velocity on the self-reported trust in the automation.  
We finally analyzed the effect of speed on each secondary task by performing an Student T-test.

\section{Results}
\label{sec:results}

\begin{figure*}[t]
	\centering
	\includegraphics[width=\textwidth]{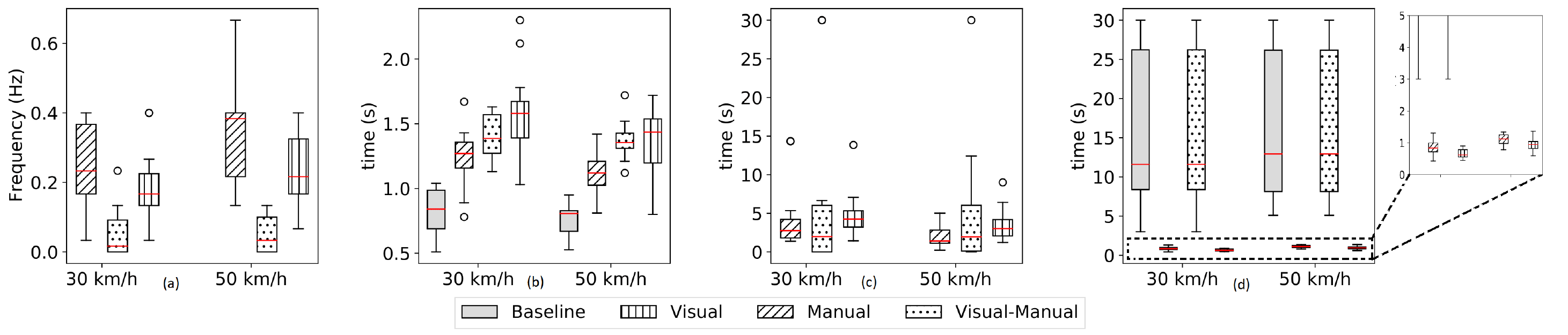}
	\caption{ Visualization of the results at the studied speeds: (a) frequency of eyes from task to road (Fr); (b) reaction time (RT); (c) average time of eyes on task (AVT) and (d) average time of eyes on road (AVR).}
	\label{fig:results:1}
\end{figure*}

\begin{figure*}[t]
    \centering
	\includegraphics[width=0.8\textwidth]{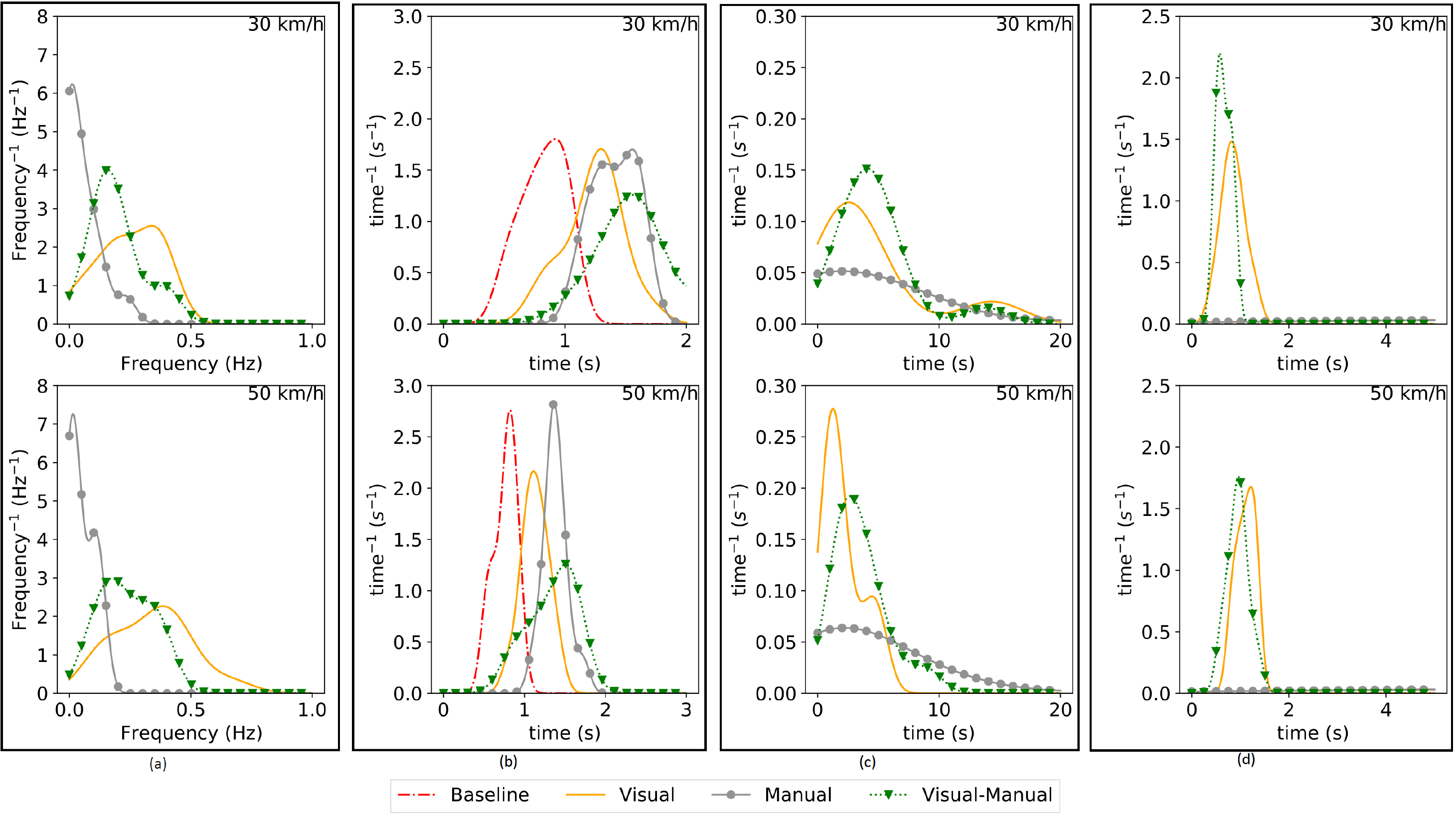}
	\caption{Visualization of the frequency distribution of each dependent variable and the baseline and secondary tasks progress at each velocity, showing (e) the dependent variable Fr; (f) RT; (g) AVT and (h) AVR. }
	\label{fig:results:2}
\end{figure*}

The results from the performed analyses at speeds of 30 and 50 km/h regarding frequency of eyes from task to road (Fr), average time of eyes on road (AVR), average time of eyes on task (AVT) and reaction times (RT) are depicted in~Figure 3. Table~\ref{table:anova} reports the results regarding the ANOVA analysis at each of the 2 velocities considered. Table~\ref{table:benferroni} presents the statistical significance obtained through post-hoc multiple comparisons to determine whether the different population means were different after applying the Bonferroni correction. Table~\ref{table:t-test} presents the statistical relationship between the dependent variables and speed for each secondary task. Finally, Table~\ref{table:wilcoxon} reports the analysis of the self-reported values regarding trust in the automated system at 30 km/h and 50 km/h and during the performance of the NDRT.


\begin{table}[t]
	\centering
	\scriptsize
	\caption{ANOVA analysis results regarding frequency of eyes from task to road (Fr), average time of eyes on road (AVR), average time of eyes on task (AVT), reaction times (RT) and perceived trust (ST). }
	\label{table:anova}
    \begin{tabular}{|p{7.65cm}|}
		\hline
		ANOVA ($\alpha$ =0.05)\\
	\end{tabular}
	\begin{tabular}{|p{1.75cm}|p{2.5cm}|p{2.5cm}|}
		\hline
		Metric&30 km/h &50 km/h\\
	\end{tabular}
	\begin{tabular}{|p{1.75cm}|p{0.8cm}|p{1.3cm}|p{0.78cm}|p{1.3cm}|}
		\hline
		&\emph{F(41)}&\emph{p}&\emph{F(41)}&\emph{p}\\
		\hline
		Fr& 13.523 & \textbf{$<$ .001}***  & 23.467 & \textbf{$<$ .001}***\\
		\hline
		AVT& 0.287 & 0.752 &1.280&0.289\\
		\hline
		AVR& 13.243 & \textbf{$<$ .001}*** & 14.462 & \textbf{$<$ .001}*** \\
		\hline
		RT&30.493&\textbf{$<$ .001}***&27.095&\textbf{$<$ .001}***\\
		\hline
		ST&1.350& 0.273 & 1.790 & 0.167 \\
		\hline
	\end{tabular}\\
\end{table}

\begin{table*}
	\centering
	\scriptsize
	\caption{Statistical analysis results from the multiple post-hoc comparison after applying the Bonferroni correction, regarding frequency of eyes from task to road (Fr), average time of eyes on road (AVR), average time of eyes on task (AVT) and reaction times (RT). From top to bottom the tables show: the mean and standard values and the comparison of the secondary tasks at the velocities 30 and 50 km. respectively }
	\label{table:benferroni}
 \begin{tabular}{|p{0.75cm}|p{3.71cm}|p{3.71cm}|p{3.705cm}|p{3.705cm}|}
  \hline
  Metric&Baseline&Manual task&Visual task& Visual-Manual task\\
 \end{tabular}\\
 \begin{tabular}{|p{0.75cm}|p{1.64cm}|p{1.63cm}|p{1.64cm}|p{1.63cm}|p{1.64cm}|p{1.63cm}|p{1.64cm}|p{1.63cm}|}
  \hline
  &30 km/h& 50 km/h&30 km/h& 50 km/h&30 km/h&50 km/h&30 km/h&50 km/h\\
 \end{tabular}\\
 \begin{tabular}{|p{.75cm}|p{0.6cm}|p{0.6cm}|p{0.6cm}|p{0.6cm}|p{0.6cm}|p{0.6cm}|p{0.6cm}|p{0.6cm}|p{0.6cm}|p{0.6cm}|p{0.6cm}|p{0.6cm}|p{0.6cm}|p{0.6cm}|p{0.6cm}|p{0.6cm}|}
  \hline
  &$M$&SD&$M$&SD&$M$&SD&$M$&SD&$M$&SD&$M$&SD&$M$&SD&$M$&SD\\
  \hline
  Fr& - & - & - & - &0.050&0.070&0.052&&0.245&0.126&0.340&0.156&0.193&0.105&0.233&0.107\\
  \hline
  AVT& - & - & - & - & 6.177& 10.367&5.039&8.024&4.351&1.172&2.088&0.421&4.727&3.051&3.537&2.136\\
  \hline
  AVR& 15.565 & 9.934 & 15.841 & 9.751 &12.529&12.518&11.693&11.200&0.852&0.239&1.104&0.188&0.662&0.148&0.949&0.212\\
  \hline
  RT&0.820&0.180&0.761&0.136&1.407&0.179&1.381&0.144&1.233&0.234&1.131&0.161&1.595&0.326&1.354&0.261\\
 \end{tabular}\\
    \begin{tabular}{|p{17.32cm}|}
  \hline
  \textbf{30 km/h}, Bonferroni-corrected p values ($\alpha$ =0.05) \\
 \end{tabular}
 \begin{tabular}{|p{.75cm}|p{2.32cm}|p{2.33cm}|p{2.32cm}|p{2.33cm}|p{2.32cm}|p{2.33cm}|}
  \hline
  Metric&Baseline vs visual task&Baseline vs manual task&Baseline vs visual-manual task& visual vs manual secondary task&visual vs visual-manual task&manual vs visual-manual task\\
 \end{tabular}
 \begin{tabular}{|p{.75cm}|p{1cm}|p{0.93cm}|p{0.95cm}|p{0.95cm}|p{0.94cm}|p{0.94cm}|p{0.95cm}|p{0.95cm}|p{0.94cm}|p{0.94cm}|p{0.95cm}|p{0.95cm}|}
  \hline
  &M diff.&\emph{p}&M diff.&\emph{p}&M diff.&\emph{p}&M diff.&\emph{p}&M diff.&\emph{p}&M diff.&\emph{p}\\
  \hline
  Fr& - & - &  - & - & - & - & 0.195 & \textbf{0.000}*** & 0.052 & 0.556 & -0.1438.714 & \textbf{0.002}**\\
  \hline
  AVT& - & - & - & - & - & - & -1.826&1.000&-0.376&1.000&1.450&1.000\\
  \hline
  AVR& 14.713 & \textbf{0.000}*** & 3.036 & 1.000 & 14.902 & \textbf{0.000}*** & -14.986& \textbf{0.000}***& 0.189 & 1.000 & 15.175& \textbf{0.000}***\\
  \hline
  RT& -0.411 & \textbf{0.000}*** & -0.586 &\textbf{0.000}***& -0.774 & \textbf{0.000}***& -0.174 &0.346& -0.362& \textbf{0.001}**&-0.188&0.247\\
    \end{tabular}
    \begin{tabular}{|p{17.32cm}|}
  \hline
  \textbf{50 km/h}, Bonferroni-corrected p values ($\alpha$ =0.05) \\
 \end{tabular}
 \begin{tabular}{|p{.75cm}|p{2.32cm}|p{2.33cm}|p{2.32cm}|p{2.33cm}|p{2.32cm}|p{2.33cm}|}
  \hline
  Metric&Baseline vs visual task&Baseline vs manual task&Baseline vs visual-manual task& visual vs manual task&visual vs visual-manual task&manual vs visual-manual task\\
 \end{tabular}
 \begin{tabular}{|p{.75cm}|p{0.94cm}|p{0.94cm}|p{0.95cm}|p{0.95cm}|p{0.94cm}|p{0.94cm}|p{0.95cm}|p{0.95cm}|p{0.94cm}|p{0.94cm}|p{0.95cm}|p{0.95cm}|}
  \hline
  &M diff.&\emph{p}&M diff.&\emph{p}&M diff.&\emph{p}&M diff.&\emph{p}&M diff.&\emph{p}&M diff.&\emph{p}\\
  \hline
  Fr& - & - &  - & - & - & - & 0.290 & \textbf{0.000}*** & 0.107 & 0.050 & 0.183 & \textbf{0.000}***\\
  \hline
  AVT& - & - & - & - & - & - & -2.951&0.3536&-1.444&1.000&1.503&1.000\\
  \hline
  AVR& 14.892 & \textbf{0.000}*** & 4.148 & 0.873 & 14.892 & \textbf{0.000}*** & 14.581& \textbf{0.000}***& 0.155 & 1.000 & 14.732& \textbf{0.000}***\\
  \hline
   RT& -3.706 & \textbf{0.000}*** & -0.370 &\textbf{0.000}***& -0.579 & \textbf{0.000}***&-0.250&\textbf{0.007}**& -0.208& \textbf{0.035}*&0.0415&1.000\\
   \hline
 \end{tabular}
 
\end{table*}

\begin{table*}
	\centering
	\scriptsize
	\caption{Results regarding the differences from the comparisons of both velocities}
	\label{table:t-test}
 \begin{tabular}{|p{10.65cm}|}
  \hline
  T-test ($\alpha$ =0.05) \textbf{30 vs . 50 km/h}\\
 \end{tabular}
 \begin{tabular}{|p{.75cm}|p{2.03cm}|p{2.04cm}|p{2.03cm}|p{2.04cm}|}
  \hline
  Metric&Baseline &Visual task&Manual task & visual-manual task\\
 \end{tabular}
 \begin{tabular}{|p{.75cm}|p{0.8cm}|p{0.8cm}|p{0.8cm}|p{0.8cm}|p{0.8cm}|p{0.8cm}|p{0.8cm}|p{0.8cm}|}
  \hline
  &\emph{t(13)}&\emph{p}&\emph{t(13)}&\emph{p}&\emph{t(13)}&\emph{p}&\emph{t(13)}&\emph{p}\\
  \hline
  Fr& - & - & -2.755 & \textbf{0.016}*& 0.000 & 1.00 & -2.264 & \textbf{0.041}*\\
  \hline
  AVT& - & - &2.197&\textbf{0.047}*&0.866&0.402&1.759&0.102\\
  \hline
  AVR& -0.165 & 0.871 & -4.303 & \textbf{0.001}** & -0.051 & 0.960 & -5.387 & \textbf{0.000}***\\
  \hline
  RT&2.391&\textbf{0.033}*&2.351&\textbf{0.035}*&0.844&0.414&3.113&\textbf{0.008}**\\
  \hline
 \end{tabular}
\end{table*}

\subsection{Frequency of eyes from task to road}

The results from the ANOVA analysis (Table~\ref{table:anova}) showed that statistically significant differences existed between the number of switches from task to road and the performed NDRT.

The multiple comparison (Table~\ref{table:benferroni}) showed that at 30 km/h there were statistically significant differences when comparing the following secondary tasks:  manual (0.050 Hz) and visual (0.246 Hz); and  manual vs. visual-manual.

At 50 km/h statistically significant differences were found between the secondary tasks manual (0.050 Hz) and visual (0.340 Hz); and manual vs visual-manual.   

Statistically significant differences additionally occurred when comparing the velocities in the visual task (0.243 Hz at 30 km/h vs 0.340 Hz at 50 km/h) and the visual-manual task (0.190 Hz at 30 km/h vs 0.233 Hz at 50 km/h). No significant statistical differences for the variable manual task were found (Table~\ref{table:t-test}).

\subsection{Average time of eyes on road}

The results from the ANOVA analysis (Table~\ref{table:anova}) showed that statistically significant differences existed between the average time that drivers were looking at the road and the performed tasks.

As shown in Table~\ref{table:benferroni} at 30 km/h, the secondary tasks in which statistically significant differences could be appreciated were baseline (15.6 s) vs visual (0.8 s), baseline vs visual-manual (0.7 s), manual (12.5 s) vs visual, and manual vs visual-manual. 
At 50 km/h  the same secondary tasks were found to have  statistically significant differences: baseline (15.8 s) vs visual (1.1 s), baseline vs visual-manual (0.9 s), manual (11.7 s) vs visual, and manual vs visual-manual. 
Comparing the eyes on road average time based on vehicle speed (velocity), statistically significant differences were assessed/found for the visual and visual-manual tasks.  No significant statistical differences for the variable manual task were found. 

\subsection{Average time of eyes on task}

At 30 km/h and 50 km/h no significant statistical differences for the  secondary tasks were found.

Results from comparing the different velocities (see (Table~\ref{table:t-test}) showed a statistically significant difference in the visual task (4.3 s at 30 km/h vs 2.0 s at 50 km/h).

\subsection{Reaction time}

The results reported in Table~\ref{table:anova} showed that statistically significant differences existed between the average reaction time to respond to a TOR and the performed tasks.

In Table~\ref{table:benferroni}, at 30 km/h there were statistically significant differences between the groups baseline condition and visual, manual and visual-manual secondary tasks, with the reaction time in the baseline scenario without secondary tasks being the fastest at 0.820 seconds (Table~\ref{table:benferroni}). 
Significant statistical differences were also found in the visual (1.23 s) and the visual-manual (1.59 s) secondary task variables. 
There were no significant statistical differences for the other variables.

The analysis of the reaction times at 50 km/h showed significant statistical differences between the visual (1.13 s) and the manual (1.38 s) secondary tasks  and between the visual and the visual-manual (1.35 s) secondary tasks. There were no significant statistical differences for the other variables visual-manual and manual.

Finally, comparing the reaction time for each secondary task at 30 vs 50 km/h respectively (Table~\ref{table:t-test}), results showed statistically significant differences in the baseline condition (0.82 s compared to 0.76 s), visual task (1.23 s vs 1.13 s) and visual-manual task (1.59 vs 1.35 s). There were no significant statistical differences for the variable manual task.

\begin{table*}
 \centering
 \scriptsize
 \caption{Statistical analysis results regarding perceived trust (ST). From top to bottom the tables show: the mean and standard values and the comparison of the secondary tasks at the velocities 30 and 50 km. respectively}
 \label{table:wilcoxon}
 \begin{tabular}{|p{0.75cm}|p{3.71cm}|p{3.71cm}|p{3.705cm}|p{3.705cm}|}
  \hline
  Metric&Baseline&Manual task&Visual task& Visual-Manual task\\
 \end{tabular}\\
 \begin{tabular}{|p{0.75cm}|p{1.64cm}|p{1.63cm}|p{1.64cm}|p{1.63cm}|p{1.64cm}|p{1.63cm}|p{1.64cm}|p{1.63cm}|}
  \hline
  &30 km/h& 50 km/h&30 km/h& 50 km/h&30 km/h&50 km/h&30 km/h&50 km/h\\
 \end{tabular}\\
 \begin{tabular}{|p{.75cm}|p{0.6cm}|p{0.6cm}|p{0.6cm}|p{0.6cm}|p{0.6cm}|p{0.6cm}|p{0.6cm}|p{0.6cm}|p{0.6cm}|p{0.6cm}|p{0.6cm}|p{0.6cm}|p{0.6cm}|p{0.6cm}|p{0.6cm}|p{0.6cm}|}
  \hline
  &$M$&SD&$M$&SD&$M$&SD&$M$&SD&$M$&SD&$M$&SD&$M$&SD&$M$&SD\\
  \hline
  ST&4.154&1.099&4.153&1.099&3.846&1.350&3.538&1.277&3.384&1.350&3.000&1.240& 2.923 &1.071 &2.692&1.202\\
 \end{tabular}\\
 \begin{tabular}{|p{17.32cm}|}
  \hline
  \textbf{30 km/h}, Wilcoxon ($\alpha$ =0.05)\\
 \end{tabular}
 \begin{tabular}{|p{.75cm}|p{2.32cm}|p{2.33cm}|p{2.32cm}|p{2.33cm}|p{2.32cm}|p{2.33cm}|}
  \hline
  Metric&Baseline vs visual task&Baseline vs manual task&Baseline vs visual-manual task& visual vs manual secondary task&visual vs visual-manual task&manual vs visual-manual task\\
 \end{tabular}
     \begin{tabular}{|p{.75cm}|p{0.98cm}|p{0.9cm}|p{0.95cm}|p{0.95cm}|p{0.94cm}|p{0.94cm}|p{0.95cm}|p{0.95cm}|p{0.94cm}|p{0.94cm}|p{0.95cm}|p{0.95cm}|}
     \hline
  &Z&\emph{p}&Z&\emph{p}&Z&\emph{p}&Z&\emph{p}&Z&\emph{p}&Z&\emph{p}\\
  \hline
  ST& -1.841 & 0.066 & -0.816 & 0.414 & -2.121 & \textbf{0.034}* & -1.300 & 0.194 & -1.134 & 0.257 & -1.414 & 0.157\\
 \end{tabular}\\
 \begin{tabular}{|p{17.32cm}|}
  \hline
  \textbf{50 km/h}, Wilcoxon ($\alpha$ =0.05)\\
 \end{tabular}
 \begin{tabular}{|p{.75cm}|p{2.32cm}|p{2.33cm}|p{2.32cm}|p{2.33cm}|p{2.32cm}|p{2.33cm}|}
  \hline
  Metric&Baseline vs visual task&Baseline vs manual task&Baseline vs visual-manual task& visual vs manual secondary task&visual vs visual-manual task&manual vs visual-manual task\\
 \end{tabular}
 \begin{tabular}{|p{.75cm}|p{0.98cm}|p{0.9cm}|p{0.95cm}|p{0.95cm}|p{0.94cm}|p{0.94cm}|p{0.95cm}|p{0.95cm}|p{0.94cm}|p{0.94cm}|p{0.95cm}|p{0.95cm}|}
     \hline
  &Z&\emph{p}&Z&\emph{p}&Z&\emph{p}&Z&\emph{p}&Z&\emph{p}&Z&\emph{p}\\
  \hline
  ST& -2.232 &\textbf{0.026}* & -2.000 & \textbf{0.046}* & -2.264 & \textbf{0.024}* & -1.732 & 0.083 & -0.447 & 0.655 & -1.300 & 0.194\\
  \hline
 \end{tabular}\\
 \begin{tabular}{|p{10.65cm}|}
  Wilcoxon ($\alpha$ =0.05) \textbf{30 vs . 50 km/h}\\
  \hline
 \end{tabular}
 \begin{tabular}{|p{.75cm}|p{2.03cm}|p{2.04cm}|p{2.03cm}|p{2.04cm}|}
  Metric&Baseline &Visual task&Manual task & visual-manual task\\
  \hline
 \end{tabular}
 \begin{tabular}{|p{.75cm}|p{0.8cm}|p{0.8cm}|p{0.8cm}|p{0.8cm}|p{0.8cm}|p{0.8cm}|p{0.8cm}|p{0.8cm}|}
     &Z&\emph{p}&Z&\emph{p}&Z&\emph{p}&Z&\emph{p}\\
     \hline
  ST& -0.577 & 0.564 & -1.089 & 0.276 & -1.732 & 0.083 & -0.828 & 0.408 \\
  \hline
 \end{tabular}\\
 
\end{table*}

\subsection{Self-reported trust in the automated driving system}

As detailed in Table~\ref{table:wilcoxon} the analysis of the self-reported values regarding trust in the automated system at 30 km/h and 50 km/h showed that there were no statistically significant differences between the levels reported and the vehicle velocity.

However, the differences between baseline and the visual, manual and  visual-manual tasks at the higher speed were statistically significant.

\section{Summary of Findings and Discussion}
\label{sec:findings}
In most cases, participants spent more time looking at the road when performing the baseline and manual task (retrieve screws from a bag) than the visual task (SCWT test). It seems that the sense of touch was sufficient to distinguish the objects. 

Similarly, participants spent more time with their gaze directed at the task during the purely visual task than during the visual-manual task (write a given text backwards). This behavior could be due to the ability of the SCWT test to demand more attentional resources. 

The increase in the vehicle velocity augmented the total time and frequency looking at the road, a lower driving speed resulting in drivers spending more time looking at their phones.

The high standard deviation that resulted from the average time of eyes on task and eyes on the road indicates the broader range of the values when performing some tasks.
As a consequence, the amount of time that drivers looked at the road depended on the task performed and the speed of the vehicle. Therefore we reject the null hypotheses defined in~\ref{subsection:ndstgb} and section~\ref{subsection:sgb} and accept the respective alternative  hypotheses \textit{H1}.   


Results showed that the reaction time to respond to a TOR was faster when no secondary tasks were performed. In the case of NDRT tasks, the SCWT test (speaking aloud a text based on a visual task) resulted in the shortest reaction time for vehicle take over. Even if the task involved a high cognitive load that is normally related to an increase of driver reaction time, the visual nature of the task made it possible to take the steering wheel quickly.

The reaction time was also shorter at a higher speed. This could be due to  a increased risk-perception at a higher velocity, that was reflected in being the participants more attentive to the road when the vehicle was faster.

Therefore, as both the vehicle speed and the secondary task performed affected driver behavior, we reject the null hypothesis defined in \ref{subsection:srt} and \ref{subsection:vrt} and accept the alternative hypotheses \textit{H1} in both cases. 

Vehicle velocity did not have an effect on the level of trust and confidence in the automated system.
Therefore, we accept the null hypothesis defined in~\ref{subsection:sst}. 

However drivers reported to be more trustful in the correct functioning of the ADS when they were not performing any other task at a higher vehicle velocity. Therefore, we reject the null hypothesis defined in~\ref{subsection:ssttask} and accept the alternative hypothesis \textit{H1}.


The short duration of segments of automated driving was necessary due to the number of conditions investigated. Therefore, even if some participants might have expected the take over request to be triggered, this applied to all of them, being the results of the experiments in this study comparable. 

An external cue in the environment that triggered the TOR did not exist. While performing the secondary tasks before the TOR was activated, the driving time was the same for everyone, ensuring the experiment's reliability.

\section{Conclusion and Future Work}
\label{sec:conclusion}

In this work we examined the impact of automation on driver behavior and analyzed the effect of vehicle speed and NDRT on the driver’s ability to regain control of a vehicle controlled by an ADS. To this end, we defined vehicle speed and NDRT as the independent variables that we modified to measure the dependent variables that related to reaction time and gaze behavior. We additionally investigated the level of confidence and trust in the automation. 

We conclude that both vehicle/driving speed and type      of secondary task affect driver behavior and, consequently, road safety as well as user trust in the automation. 

Future work will address extended scenarios with a variety of road geometry and obstacles ahead as well as external cues to trigger the TOR. We aim to explore additional velocities and will compare the findings with simulated scenarios.

\section*{ACKNOWLEDGMENT}
This work was supported by the Austrian Ministry for Climate Action, Environment, Energy, Mobility, Innovation and Technology (BMK) Endowed Professorship for Sustainable Transport Logistics 4.0.

\bibliographystyle{IEEEtran}
\bibliography{paper}

\begin{thebibliography}{10}
\providecommand{\url}[1]{#1}
\csname url@samestyle\endcsname
\providecommand{\newblock}{\relax}
\providecommand{\bibinfo}[2]{#2}
\providecommand{\BIBentrySTDinterwordspacing}{\spaceskip=0pt\relax}
\providecommand{\BIBentryALTinterwordstretchfactor}{4}
\providecommand{\BIBentryALTinterwordspacing}{\spaceskip=\fontdimen2\font plus
\BIBentryALTinterwordstretchfactor\fontdimen3\font minus
  \fontdimen4\font\relax}
\providecommand{\BIBforeignlanguage}[2]{{%
\expandafter\ifx\csname l@#1\endcsname\relax
\typeout{** WARNING: IEEEtran.bst: No hyphenation pattern has been}%
\typeout{** loaded for the language `#1'. Using the pattern for}%
\typeout{** the default language instead.}%
\else
\language=\csname l@#1\endcsname
\fi
#2}}
\providecommand{\BIBdecl}{\relax}
\BIBdecl

\bibitem{olaverri2016human}
C.~Olaverri-Monreal and T.~Jizba, ``{Human factors in the design of
  human--machine interaction: An overview emphasizing V2X communication},''
  \emph{IEEE Transactions on Intelligent Vehicles}, vol.~1, no.~4, pp.
  302--313, 2016.

\bibitem{olaverri2016autonomous}
C.~Olaverri-Monreal, ``{Autonomous vehicles and smart mobility related
  technologies},'' \emph{Infocommunications Journal}, vol.~8, no.~2, pp.
  17--24, 2016.

\bibitem{olaverri2017road}
------, ``Road safety: Human factors aspects of intelligent vehicle
  technologies,'' in \emph{Smart Cities, Green Technologies, and Intelligent
  Transport Systems}.\hskip 1em plus 0.5em minus 0.4em\relax Springer, 2017,
  pp. 318--332.

\bibitem{Olaverri-Monreal2020}
------, ``Promoting trust in self-driving vehicles,'' \emph{Nature
  Electronics}, vol.~3, no.~6, p. 292–294, Jun 2020.

\bibitem{morales2020automated}
W.~Morales-Alvarez, O.~Sipele, R.~L{\'e}beron, H.~H. Tadjine, and
  C.~Olaverri-Monreal, ``Automated driving: A literature review of the take
  over request in conditional automation,'' \emph{Electronics}, vol.~9, no.~12,
  p. 2087, 2020.

\bibitem{ntsb2020}
\BIBentryALTinterwordspacing
{National Transportation Safety Board Office of Safety Recommendations and
  Communications}, ``{Tesla Crash Investigation Yields 9 NTSB Safety
  Recommendations}.'' [Online]. Available:
  \url{https://www.ntsb.gov/news/press-releases/Pages/NR20200225.aspx}
\BIBentrySTDinterwordspacing

\bibitem{RUSCIO2017105}
\BIBentryALTinterwordspacing
D.~Ruscio, A.~Bos, and M.~Ciceri, ``Distraction or cognitive overload? using
  modulations of the autonomic nervous system to discriminate the possible
  negative effects of advanced assistance system,'' \emph{Accident Analysis \&
  Prevention}, vol. 103, pp. 105 -- 111, 2017. [Online]. Available:
  \url{http://www.sciencedirect.com/science/article/pii/S0001457517301276}
\BIBentrySTDinterwordspacing

\bibitem{Blatt1994}
\BIBentryALTinterwordspacing
S.~J. Blatt and R.~Q. Ford, \emph{Theoretical and Methodological Issues in the
  Study of Therapeutic Change}.\hskip 1em plus 0.5em minus 0.4em\relax Boston,
  MA: Springer US, 1994, pp. 1--27. [Online]. Available:
  \url{https://doi.org/10.1007/978-1-4899-1010-3_1}
\BIBentrySTDinterwordspacing

\bibitem{Gold2016TakingSituations}
\BIBentryALTinterwordspacing
C.~Gold, M.~K{\"{o}}rber, D.~Lechner, and K.~Bengler, ``{Taking Over Control
  From Highly Automated Vehicles in Complex Traffic Situations},'' \emph{Human
  Factors: The Journal of the Human Factors and Ergonomics Society}, vol.~58,
  no.~4, pp. 642--652, 6 2016. [Online]. Available:
  \url{http://journals.sagepub.com/doi/10.1177/0018720816634226}
\BIBentrySTDinterwordspacing

\bibitem{Radlmayr2014}
\BIBentryALTinterwordspacing
J.~Radlmayr, C.~Gold, L.~Lorenz, M.~Farid, and K.~Bengler, ``{How Traffic
  Situations and Non-Driving Related Tasks Affect the Take-Over Quality in
  Highly Automated Driving},'' \emph{Proceedings of the Human Factors and
  Ergonomics Society Annual Meeting}, vol.~58, no.~1, pp. 2063--2067, 9 2014.
  [Online]. Available:
  \url{http://journals.sagepub.com/doi/10.1177/1541931214581434}
\BIBentrySTDinterwordspacing

\bibitem{8594655}
A.~{Eriksson}, S.~M. {Petermeijer}, M.~{Zimmermann}, J.~C.~F. {de Winter},
  K.~J. {Bengler}, and N.~A. {Stanton}, ``Rolling out the red (and green)
  carpet: Supporting driver decision making in automation-to-manual
  transitions,'' \emph{IEEE Transactions on Human-Machine Systems}, vol.~49,
  no.~1, pp. 20--31, 2019.

\bibitem{gold2018modeling}
C.~Gold, R.~Happee, and K.~Bengler, ``Modeling take-over performance in level 3
  conditionally automated vehicles,'' \emph{Accident Analysis \& Prevention},
  vol. 116, pp. 3--13, 2018.

\bibitem{du2020evaluating}
N.~Du, J.~Kim, F.~Zhou, E.~Pulver, D.~M. Tilbury, L.~P. Robert, A.~K. Pradhan,
  and X.~J. Yang, ``Evaluating effects of cognitive load, takeover request lead
  time, and traffic density on drivers’ takeover performance in conditionally
  automated driving,'' in \emph{12th International Conference on Automotive
  User Interfaces and Interactive Vehicular Applications}, 2020, pp. 66--73.

\bibitem{Martens1997THEEO}
M.~H. Martens, S.~L. Compte, and N.~A. Kaptein, ``{The Effects Of Road Design
  On Speed Behaviour: A Literature Review}.''\hskip 1em plus 0.5em minus
  0.4em\relax Deliverable D1 (Report 2.3.1), Managing Speed on European Roads
  (MASTER) project. VTT. Finland, 1997.

\bibitem{9294751}
W.~{Morales-Alvarez}, N.~{Smirnov}, E.~{Matthes}, and C.~{Olaverri-Monreal},
  ``Vehicle automation field test: Impact on driver behavior and trust,'' in
  \emph{2020 IEEE 23rd International Conference on Intelligent Transportation
  Systems (ITSC)}, 2020, pp. 1--6.

\bibitem{Neukum2014}
F.~Naujoks, C.~Mai, and A.~Neukum, ``{The effect of urgency of take-over
  requests during highly automated driving under distraction conditions},'' in
  \emph{5th International Conference on Applied Human Factors and Ergonomics
  AHFE}, 2014.

\bibitem{Brandenburg2019Take-overConditions}
S.~Brandenburg and L.~Chuang, ``{Take-over requests during highly automated
  driving: How should they be presented and under what conditions?}''
  \emph{Transportation Research Part F: Traffic Psychology and Behaviour},
  vol.~66, pp. 214--225, 10 2019.

\bibitem{Borojeni2018FeelVehicles}
\BIBentryALTinterwordspacing
S.~S. Borojeni, S.~C. Boll, W.~Heuten, H.~H. B{\"{u}}lthoff, and L.~Chuang,
  ``{Feel the movement: Real motion influences responses to Take-over requests
  in highly automated vehicles},'' in \emph{Conference on Human Factors in
  Computing Systems - Proceedings}, vol. 2018-April.\hskip 1em plus 0.5em minus
  0.4em\relax New York, New York, USA: Association for Computing Machinery, 4
  2018, pp. 1--13. [Online]. Available:
  \url{http://dl.acm.org/citation.cfm?doid=3173574.3173820}
\BIBentrySTDinterwordspacing

\bibitem{radlmayr2014a}
J.~Radlmayr, C.~Gold, L.~Lorenz, M.~Farid, and K.~Bengler, ``How traffic
  situations and non-driving related tasks affect the take-over quality in
  highly automated driving,'' in \emph{Proceedings of the human factors and
  ergonomics society annual meeting}, vol.~58, no.~1.\hskip 1em plus 0.5em
  minus 0.4em\relax Sage Publications Sage CA: Los Angeles, CA, 2014, pp.
  2063--2067.

\bibitem{zeeb2016take}
K.~Zeeb, A.~Buchner, and M.~Schrauf, ``Is take-over time all that matters? the
  impact of visual-cognitive load on driver take-over quality after
  conditionally automated driving,'' \emph{Accident Analysis \& Prevention},
  vol.~92, pp. 230--239, 2016.

\bibitem{wandtner2018effects}
B.~Wandtner, N.~Sch{\"o}mig, and G.~Schmidt, ``Effects of non-driving related
  task modalities on takeover performance in highly automated driving,''
  \emph{Human factors}, vol.~60, no.~6, pp. 870--881, 2018.

\bibitem{gold2013take}
C.~Gold, D.~Damb{\"o}ck, L.~Lorenz, and K.~Bengler, ``“take over!” how long
  does it take to get the driver back into the loop?'' in \emph{Proceedings of
  the human factors and ergonomics society annual meeting}, vol.~57,
  no.~1.\hskip 1em plus 0.5em minus 0.4em\relax Sage Publications Sage CA: Los
  Angeles, CA, 2013, pp. 1938--1942.

\bibitem{Eriksson2017}
\BIBentryALTinterwordspacing
A.~Eriksson and N.~A. Stanton, ``{Takeover Time in Highly Automated Vehicles:
  Noncritical Transitions to and from Manual Control},'' \emph{Human Factors},
  vol.~59, no.~4, pp. 689--705, 6 2017. [Online]. Available:
  \url{http://journals.sagepub.com/doi/10.1177/0018720816685832}
\BIBentrySTDinterwordspacing

\bibitem{allamehzadeh2016automatic}
A.~Allamehzadeh and C.~Olaverri-Monreal, ``Automatic and manual driving
  paradigms: Cost-efficient mobile application for the assessment of driver
  inattentiveness and detection of road conditions,'' in \emph{2016 IEEE
  Intelligent Vehicles Symposium (IV)}.\hskip 1em plus 0.5em minus 0.4em\relax
  IEEE, 2016, pp. 26--31.

\bibitem{verberne2012trust}
F.~M. Verberne, J.~Ham, and C.~J. Midden, ``Trust in smart systems: Sharing
  driving goals and giving information to increase trustworthiness and
  acceptability of smart systems in cars,'' \emph{Human factors}, vol.~54,
  no.~5, pp. 799--810, 2012.

\bibitem{olaverri2014capturing}
C.~Olaverri-Monreal and J.~Gon{\c{c}}alves, ``Capturing mental models to meet
  users expectations,'' in \emph{2014 9th Iberian Conference on Information
  Systems and Technologies (CISTI)}.\hskip 1em plus 0.5em minus 0.4em\relax
  IEEE, 2014, pp. 1--5.

\bibitem{hussein2016p2v}
A.~Hussein, F.~Garcia, J.~M. Armingol, and C.~Olaverri-Monreal, ``P2v and v2p
  communication for pedestrian warning on the basis of autonomous vehicles,''
  in \emph{2016 IEEE 19th International Conference on Intelligent
  Transportation Systems (ITSC)}.\hskip 1em plus 0.5em minus 0.4em\relax IEEE,
  2016, pp. 2034--2039.

\bibitem{petersen2019situational}
L.~Petersen, L.~Robert, J.~Yang, and D.~Tilbury, ``Situational awareness,
  driver’s trust in automated driving systems and secondary task
  performance,'' \emph{SAE International Journal of Connected and Autonomous
  Vehicles, Forthcoming}, 2019.

\bibitem{beller2013improving}
J.~Beller, M.~Heesen, and M.~Vollrath, ``Improving the driver--automation
  interaction: An approach using automation uncertainty,'' \emph{Human
  factors}, vol.~55, no.~6, pp. 1130--1141, 2013.

\bibitem{capalar2017hypovigilance}
J.~{\c{C}}apalar and C.~Olaverri-Monreal, ``Hypovigilance in limited
  self-driving automation: Peripheral visual stimulus for a balanced level of
  automation and cognitive workload,'' in \emph{2017 IEEE 20th International
  Conference on Intelligent Transportation Systems (ITSC)}.\hskip 1em plus
  0.5em minus 0.4em\relax IEEE, 2017, pp. 27--31.

\bibitem{xiong2012use}
H.~Xiong, L.~N. Boyle, J.~Moeckli, B.~R. Dow, and T.~L. Brown, ``Use patterns
  among early adopters of adaptive cruise control,'' \emph{Human factors},
  vol.~54, no.~5, pp. 722--733, 2012.

\bibitem{openpilot}
\BIBentryALTinterwordspacing
{Coma.AI}, ``{commaai\//openpilot}.'' [Online]. Available:
  \url{https://github.com/commaai/openpilot}
\BIBentrySTDinterwordspacing

\bibitem{vps}
\BIBentryALTinterwordspacing
V.~{G}mb{H}, ``{Eye Hyper-Tracking. See what others see - Viewpointsystem
  Eye-Tracking}.'' [Online]. Available: \url{https://viewpointsystem.com/en/}
\BIBentrySTDinterwordspacing

\bibitem{jensen1966stroop}
A.~R. Jensen and W.~D. Rohwer~Jr, ``The stroop color-word test: a review,''
  \emph{Acta psychologica}, vol.~25, pp. 36--93, 1966.

\end{thebibliography}
\end{document}